# Evaluation of different rectangular scan strategies for STEM imaging

A. Velazco, M. Nord, A. Béché, J. Verbeeck

EMAT, University of Antwerp, Groenenborgerlaan 171, 2020 Antwerp, Belgium

Abstract

STEM imaging is typically performed by raster scanning a focused electron probe over a sample. Here we investigate and compare three different scan patterns, making use of a programmable scan engine that allows to arbitrarily set the sequence of probe positions that are consecutively visited on the sample. We compare the typical raster scan with a so-called 'snake' pattern where the scan direction is reversed after each row and a novel Hilbert scan pattern that changes scan direction rapidly and provides an homogeneous treatment of both scan directions. We experimentally evaluate the imaging performance on a single crystal test sample by varying dwell time and evaluating behaviour with respect to sample drift. We demonstrate the ability of the Hilbert scan pattern to more faithfully represent the high frequency content of the image in the presence of sample drift. It is also shown that Hilbert scanning provides reduced bias when measuring lattice parameters from the obtained scanned images while maintaining similar precision in both scan directions which is especially important when e.g. performing strain analysis. Compared to raster scanning with flyback correction, both snake and Hilbert scanning benefit from dose reduction as only small probe movement steps occur.

**Keywords:** Aberration corrected STEM, scanning distortions, drift, precision, Hilbert scan pattern

## 1. Introduction

Scanning transmission electron microscopes (STEMs) equipped with spherical aberration correctors are well-known for their ability to obtain images with atomic scale information [1,2]. In a STEM, multiple signals such as incoherent annular dark field, coherent bright field, analytical, etc., can be acquired simultaneously while scanning a focused electron probe consecutively across the sample. Along with the development of aberration correctors, improvements in the microscope environment (vibration damping, precise air conditioning and water cooling systems, acoustic and electromagnetic shielding, etc.) have also been key to achieve sub-Ångstrom resolution. Because of the inherent serial nature of the scanning, external and internal factors creating probe and sample instabilities can have a detrimental effect on the image quality. Currently, such factors cannot be completely countered and environmental disturbances can create unwanted imaging artefacts in even the most advanced STEM instruments [3,4,5]. As in many other scanning probe microscopy (SPM) techniques, the conventional scanning path adopted for STEM imaging is raster scanning. Here the probe is deflected from left to right in horizontal lines, and then shifting the probe position vertically for the next line. The measured signals are digitized and represented as 2D arrays of pixels in quasi real-time. Intrinsically, this common scan strategy is prone to image distortions due to the following reasons.

The probe is scanned faster in the horizontal direction compared to the vertical direction, resulting in an unequal treatment of the x and y directions in the image. When the probe reaches the end of one line it is deflected back to the beginning of the next line in the so-called 'flyback' motion. The finite time response of the scan coil system can lead to a deviation between the actual and the targeted position of the probe during this flyback. This results in a deformation of the first few pixels of a new scan line until a constant probe velocity is reached. This effect is typically alleviated by adding a time delay (flyback time) at the beginning of each line. Typical flyback delays may vary from 100-2000 µs depending on the details of the microscope and it could be necessary to adjust this according to the acquisition conditions such as magnification and dwell time (note that even with flyback time, distortion of the first few scan positions might be present as constant velocity is still building up). Besides increasing the total acquisition time, this solution has the drawback of injecting electron dose into the flyback position which is not used for imaging if the beam cannot be shuttered during that time. Such increased dose and unequal distribution can lead to issues with beam sensitive samples degrading in ways that are difficult to predict.

When the scan lines are misaligned [6], distortions occur that lowers the accuracy and precision with which parameters can be obtained from images in a quantitative analysis.

On top of these scanning artefacts [6,7,8], sample drift is common and results in an unequal effect on the fast versus the slow scan direction. This leads to distortions in the image, which may be subtle to notice but might prevent crystallographic analysis. These distortions become far more obvious in a diffractogram, calculated by Fourier transforming images which contain periodic lattice features. Such diffractograms then typically show streaks in the vertical (slow scan) direction as well as extra modulation peaks stemming from unwanted periodic modulation of the probe position (e.g. due to stray fields). A skewed lattice image could be representative of the structure or the result of drift, more complex non-linear drift could create curved atomic fringes in the images [9]. In general, the slow scan direction is more prone to this kind of distortions.

A considerable number of theoretical models and algorithms has been proposed to address distortions in images acquired with raster scanning, such as linear drift correction based on a prior knowledge of the crystal structure [7], scanning artifacts and drift correction tackled by non-rigid registration of a series of images [10], and ultimately corrections of scanning instabilities and non-linear drift from only a pair of images acquired with orthogonal scan directions [6,9]. However, in STEM only a modest number of investigations focused on alternative scanning paths to reduce these distortions. Spiral scanning paths were investigated by Sang et al. [11] as a method to acquire images at extremely high speed without the need of flyback delay and with a near single harmonic frequency spectrum of the scan coil drive signals. Different spirals scans were investigated and classified as constant angular velocity and constant linear velocity spirals, but both suffer from non-uniform image quality over the scanned area. In the first case, to keep the angular velocity constant, the sampling density and applied dose on the edges is lower than in the central region. In the second case, to keep the linear velocity constant and the sampling density and applied dose uniform, the beam moves faster in the central region than on the edges possibly creating image distortions as the scan system reacts differently to the changing frequency that is applied. A post-processing step is needed to correct for this distortion.

Another example is given by L. Kovarik et al. [12] for sparse acquisition in STEM. They reduce distortions caused by the finite response time of the scan system by applying a line-hopping scanning procedure where the speed of the probe is constant in the fast scan direction and the beam moves randomly across the slow scan direction.

An alternative scanning method that avoids big flyback jumps and delays, improves on drift induced distortions and requires no post-processing would be attractive. The Hilbert space filling curve [13] has been proposed previously in the field of scanning probe microscopy (SPM) [14] as a rectangular scanning alternative to the raster method. The Hilbert scanning method changes the direction of the scanning every one or two steps, resulting in only small jumps during scanning, eliminating the slow scan direction and making the path more isotropic, while each point is still scanned exactly once as in the raster method. In this paper we will investigate the Hilbert scanning method and compare it to raster scanning in terms of distortions on the shape of atomic columns and in terms of lattice distortions. We will show that both are complementary means to measure image quality depending on the specific research goal. For comparison, a snake scanning method, which changes the scan direction for each scan line [11], eliminating big flyback jumps as well, was also tested.

## 2. Experimental

Experiments were carried out on a probe aberration corrected FEI TITAN microscope, operated at 300 kV, with 50 pA of beam current and a spatial resolution of approximately 0.8 Å. A reference sample of strain-free $SrTiO_3$ (STO) was prepared by FIB in [001] zone axis orientation.

A custom hardware scan engine [15] was employed to control the scan inputs of the microscope and to acquire the high-angle annular dark-field (HAADF) signal which was progressively displayed in a 2D array according to the scanning sequence. The scanning sequence could be freely altered from a hardware look up table in the scan engine. Different scanning patterns were generated with a simple Matlab script that provided a series of x- and y-coordinates ordered in the sequence of scanning.

The three scanning methods employed here are illustrated in Figure 1, and are labeled raster, snake and Hilbert scanning. Blue dots indicate the scanning positions and the black arrows the scanning path. The snake and Hilbert scanning eliminate the need for flyback delays while the pixel size is kept constant for those three cases.

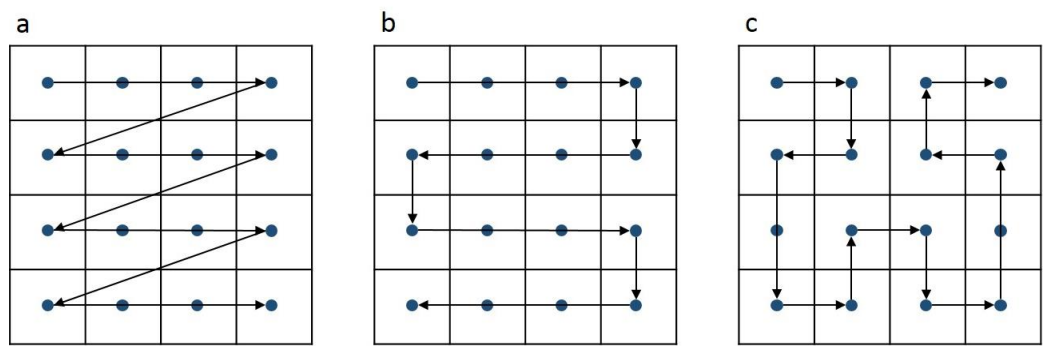

Figure 1. Different scanning methods. The blue dots indicate the scanning positions and the black arrows the scanning path. (a) Raster scanning, (b) snake scanning and (c) Hilbert scanning of order 2. A Hilbert curve of order n is created recursively by repeating four times the same pattern of order n-1. Only for the raster scanning method a flyback delay was necessary to reduce distortions because of the finite settling time of the probe steering system.

For each dwell time three images, at the same magnification, with a frame size of 1024 × 1024 pixels were acquired using the three scanning methods described above. A series of images were acquired at 2, 4, 7, 8, 9, 10, 12, 15 and 20 µs dwell time, representative of most experimental STEM imaging conditions. The entire experiment was done over the same area of the sample with the Sr sublattice (010) planes oriented close to the fast scan direction and the (100) planes oriented close to the slow scan direction, indicated in Figure 2(a). Only for the raster scanning a flyback delay was necessary. The delay was set to 1 ms, which worked well for all the raster images acquired at different dwell times. This delay added an extra acquisition time per image of approximately 1 second, giving an extra electron dose of approximately $6x10^8$ $e^-/Å^2$ per image (images acquired with the raster method without flyback delay can be found in the Supporting Information Figure s1, distortions on the left side of the images are easy to identify). Moderate random sample drift was observed during the whole duration of the experiment.

## 3. Results

In Figure 2, experimental images acquired at 15 µs dwell time are shown. Figure 2(a1) was acquired with raster scanning, the diffractogram shows vertical streaks in the slow scan direction (faint extra spots are also visible when using raster scanning at higher dwell time, see Supporting Information Figure s2). The vertical streaks are a manifestation of non-periodic distortions that arise from drift of the sample or scan instabilities that in combination with the periodicity of the line by line scanning results in modulation streaks in the slow scan direction. The dephasing (shifting) between the individual fast scan lines can be understood as a random shifting of the lines in the image depending on their vertical scan position. Figure 2(b1) was acquired with snake scanning, in addition to the vertical streaks, the diffractogram shows extra spots present on top of the streaks. The extra spots are the result of the distortions induced by the horizontal line shifting which, compared to the raster scanning, is no longer purely random. Indeed, every scan line will have phase continuity at the left or right hand side of the image. More details regarding the origin of the shift are given further. This causes the modulation streaks to have correlation peaks at Nyquist frequency in the slow scan direction. An investigation of the snake scanning method in scanning electron microscopy (SEM) with sub-pixel shift correction of the line shifting by digital image correlation and phase correlation was reported by C. Lenthe William et al [16]. Figure 2(c1) was acquired with Hilbert scanning, the diffractogram shows that vertical streaks are absent and only some extra spots at high frequency are visible in the vertical and horizontal direction. In the Hilbert scanning method, the scanning direction of the probe is changed every few steps, creating a shift between two short scan lines, similar to the snake scanning case. The short scan lines can be oriented in the vertical or horizontal direction, introducing the extra spots in the diffractogram, but now far less obvious compared to the snake scanning case due to the irregular pattern.

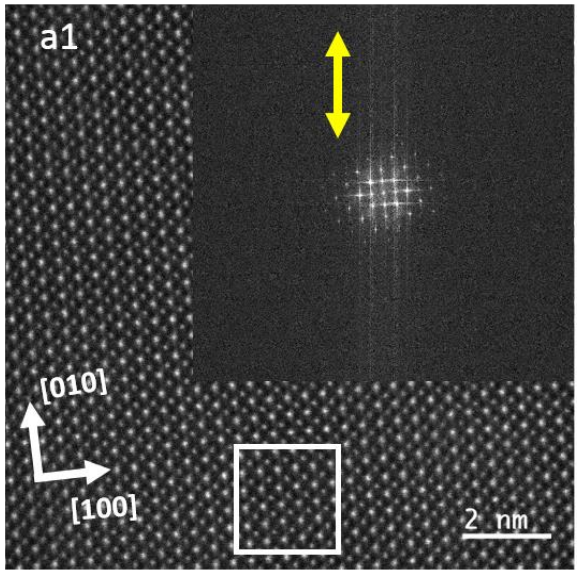 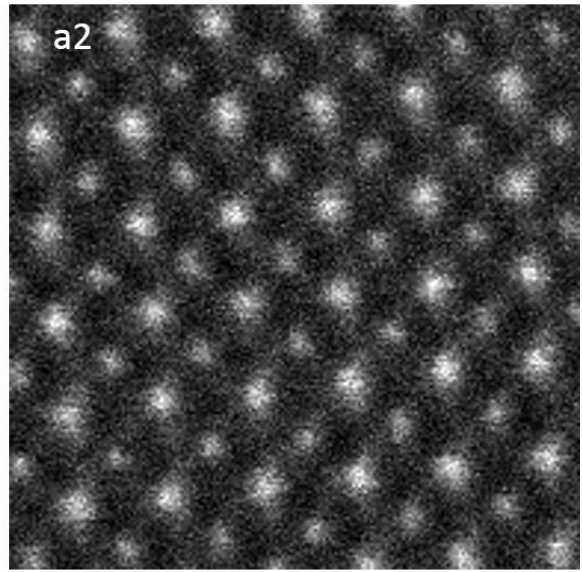
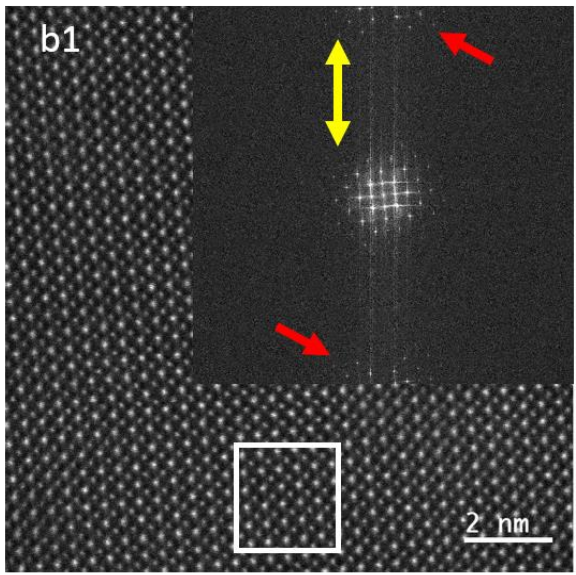 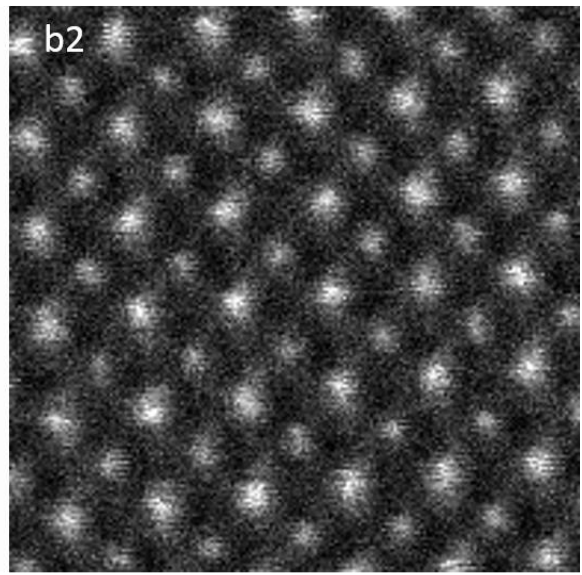
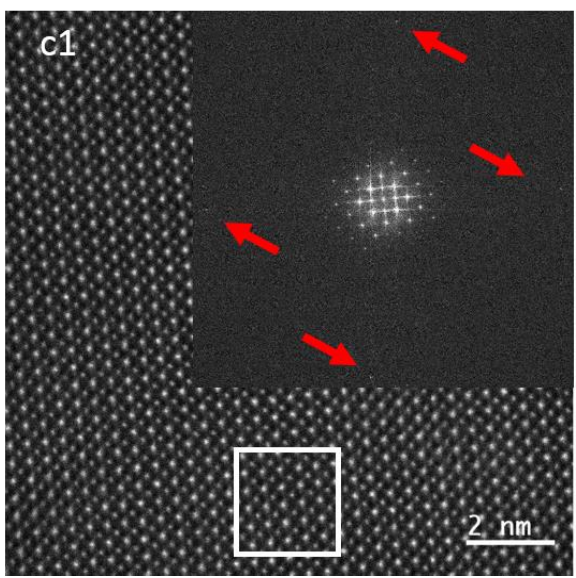 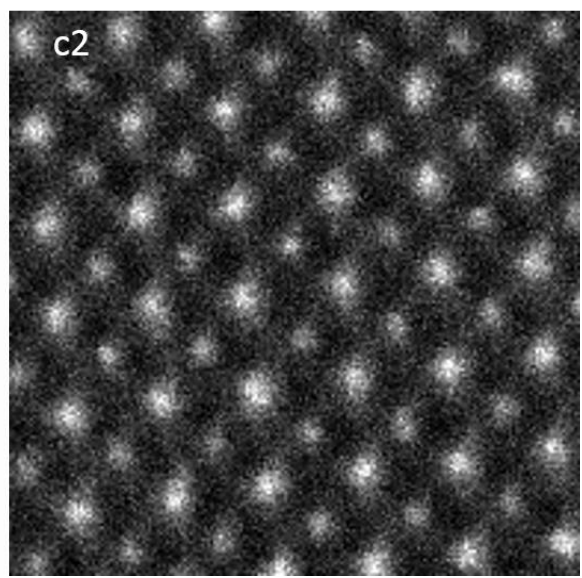

Figure 2. Experimental HAADF images acquired at 15 µs dwell time. (a1) Image acquired with raster scanning. Calculated diffractogram in the inset, the vertical streaks, indicated by yellow double arrows, are manifestations of drift or random external influences between scan lines. (a2) Highlighted area in (a1). (b1) Image acquired with snake scanning. Calculated diffractogram in the inset, in addition to the vertical streaks, indicated by yellow double arrows, some extra spots indicated by red single arrows are visible forming a replica of the central frequencies at Nyquist. The extra spots are the result of the periodic distortions between scan lines with periodicity of 2 pixels. (b2) Highlighted area in (b1). (c1) Image acquired with Hilbert scanning. Calculated diffractogram in the inset, non-vertical streaks are visible. Weak extra spots in the horizontal and vertical direction are indicated by the red single arrows. The extra spots are the result of the inversion of the scanning direction after a short number of steps, creating a shift between two short horizontal or vertical scan lines, similar to the snake scanning case. (c2) Highlighted area in (c1).

Distortions along the slow scan direction are easily identified as torn atomic columns in the image acquired with the snake scanning, seen in the enlarged image in Figure 2(b2). Those distortions are less noticeable in the images acquired with the raster and Hilbert scanning, in that specific order, see enlarged images in Figures 2(a2) and 2(c2), respectively. The observed shifts are caused predominantly by a time delay between the scan generator and the reading of the HAADF signal, the bandwidth limited amplifiers driving the scan coils, the bandwidth limitation in the detector amplifier and to a much lesser extent by bandwidth limitations in the scan generator. This was verified by feeding the output of the scan generator directly to its input, resulting in negligible time effects in the applied dwell time regime. A delay of less than a dwell time between the output and the input of the scan generator was measured for all the dwell times we employed here.

Any delay or the settling time effect on probe positioning and possibly detector signal is less visible in raster scanning as it simply shifts the whole image (assuming adequate flyback time and avoiding the remaining distortions in the first few pixels of a row). In snake scanning it results in a similar shift but now with reversed direction for odd and even rows, making it more noticeable compared to the raster scanning. In Hilbert scanning, the settling time effect is most critical as the scan direction is regularly changing.

To alleviate this, we model the relation between target scan positions and the actual probe positions which lags in time with a Gaussian impulse response that is given as:

$$h(t) = \exp(-(t - t_d)^2/\sigma^2) \qquad (1)$$

With $t_d$ representing the time delay of the entire setup and $\sigma$ chosen constant with a value of 1/8 of the corresponding dwell time for numerical reasons. Convolving the x and y scan signals with this impulse response function allows us to estimate the actual probe position from the target positions with sub pixel accuracy, taking into account the effect of the followed path.

We get for the actual positions in x and y as a function of time:

$$x(t) = x\_target(t) * h(t); y(t) = y\_target(t) * h(t) \qquad (2)$$

where $*$ denotes convolution. We optimised $t_d$ in order to minimize the undesired reflections in the diffractogram for each experiment. For the snake scanning case, at dwell times longer than 4 µs, this parameter was less than a dwell time; these values were similar to the values measured feeding the output of the scan generator directly to its input. At 2 and 4 µs dwell time these parameters were approximately 1.4 and 2.5 times a dwell time, respectively, which shows the increasing time lag of scan system and detector when dwell times become shorter. For the Hilbert scanning case, the delays were similar to the scan engine alone for 12, 15 and 20 µs dwell time. For acquisitions faster than 12 µs the delay obtained from our model was approximately between 1.8 to 3 times a dwell time. This was expected as in this case the probe is constantly changing scanning direction and it cannot reach a constant speed as in the snake scanning case (an image showing the target positions, the scanning path as measured from the scan generator and the model path used to correct the Hilbert image acquired at 15 μs of dwell time can be found in the Supporting Information Figure s3).

We use these estimated probe positions to correct the experimental images by using linear interpolation of the image intensities to the intended rectangular target pixel grid (the calculated diffractograms of the snake and Hilbert images acquired at 15 μs of dwell time and the ones of their corresponding corrected images are shown in the Supporting Information Figure s4. For the snake scanning, most of the extra spots at high frequencies disappeared while the vertical streaks remain. For the Hilbert scanning, most of the extra spots at high frequencies disappeared). This correction treats all methods with exactly the same algorithm and avoids up and downsampling schemes which can alter the frequency content of the image, and make it harder to objectively compare methods due to unintended filtering and smoothing. In the remainder of this paper we will report both corrected and uncorrected datasets for the snake and Hilbert scanning types. It has to be noted that the assumption of this impulse response is likely an oversimplification, but as long as dwell times are not chosen too short the correction is very minor. Moreover, a more accurate modeling of the time response of the scan and detector system would require a detailed system analysis which would anyway be different for different microscope models and vendors and can even depend on magnification, beam current and dwell time settings according to the internal design choices that were made in the microscope hardware.

To compare and quantify the distortions when scanning with the different methods, images were processed with the help of the Atomap software [17]. First distortions in terms of atomic column shape were quantified, followed by a study of observed lattice distortions. For the distortions of the atomic columns shape, 2-D Gaussian functions were fitted to the atomic columns on the Sr sublattice to estimate their positions. More information on how these is calculated can be found in the work reported by M. Nord et al. [17]. The region around the atomic columns were masked and for each of the masked images the center of mass of every row was calculated. The centers of mass were compared to a linear fitting (a curve oriented close to the perpendicular direction of the rows, to account for drift, misalignment, astigmatism, or anything that would cause the atoms to deviate from a circularly symmetric intensity pattern) and the standard deviation from this line was calculated to represent the local distortions in the slow scan direction[1]. The same procedure was applied to the centers of mass of the columns on the masked images to represent the distortions in the fast scan direction. The root-mean-square (RMS) of both the standard deviations in the fast and slow scan direction corresponding to every atomic column, $\sigma_{RMSx}$ and $\sigma_{RMSy}$ respectively, were calculated. The same methodology was applied to every image acquired with the different scanning methods at different dwell times. These deviations quantify the high frequency noise content as this noise is commonly measured from lines profiles along lattice planes [7].

The results are plotted in Figure 3, including the results from the uncorrected images acquired with the snake and Hilbert scanning and their corresponding corrected images. As expected scanning with the snake method generates more deviations in the y direction (slow scan direction). The data from the snake corrected images shows a reduction of the distortions in the y direction; however, a reduction of the distortions in the x direction is also present. The Hilbert scanning method, as expected, generates more symmetric deviations in x and y. For acquisitions with dwell time longer than 4 μs, the deviations in the y direction are smaller compared with the raster scanning method. For the uncorrected Hilbert scanning data, compared to raster scanning, a reduction of the deviations in the y direction of up to 21.5% was achieved while the deviations in the x direction are higher with an increase of up to 21%. The lowest deviation produced when scanning with the Hilbert method can be identified as the data point closer to the origin, which was at 10 μs dwell time. In Figure 3 the dotted curves correspond to circles of equal root mean squared total deviation. For the corrected Hilbert scanning data, a significant reduction of the deviations in x and y direction is depicted in the figure showing that time lag correction is very important here. Compared to the raster scanning data, now the deviations in the y direction are smaller for all the dwell times.

---

[1] For the software implementation of this algorithm, see
https://atomap.org/quantify_scan_distortions

The total deviation, calculated as the RMS of $\sigma_{RMSx}$ and $\sigma_{RMSy}$, for all the scanning conditions and their corresponding corrections are given in Table 1. For a dwell time longer than 4 μs, the total deviations generated with the Hilbert method (data calculated from the uncorrected images) are comparable to the ones generated with the raster method with the lowest deviation generated at 10 μs dwell time. The total deviation corresponding to the corrected Hilbert data is less than the total deviation calculated for all the other cases. This shows that as far as the image of the atoms is concerned, hilbert scanning provides clear benefits in terms of scan distortions and equal treatment of both scan directions without requiring e.g. a double acquisition with 90 degree rotated scan direction as is often done for raster scanning.

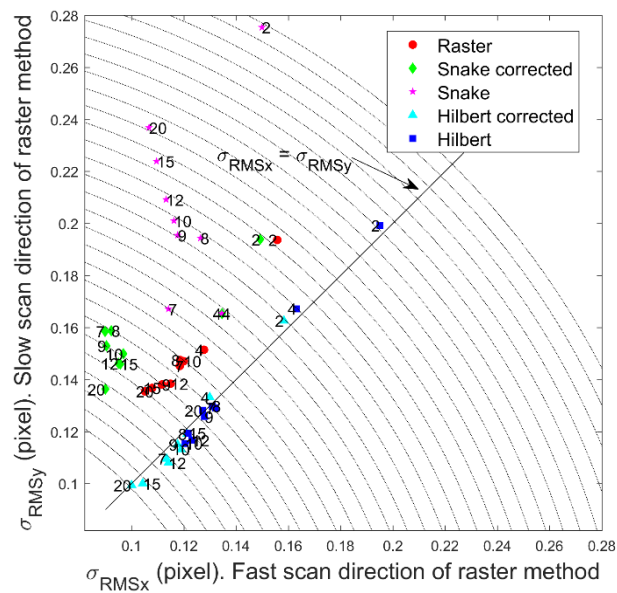

Figure 3. Measured standard deviation of the location of the center of mass of the intensity profile of line scans through the atomic columns in the two perpendicular directions, $\sigma_{RMSx}$ and $\sigma_{RMSy}$ for the different scanning methods and dwell times. Dwell times are given in μs and are indicated by the numbers next to each data point. The dotted curves correspond to circles of constant RMS deviation in both x and y: $cte = \sqrt{\sigma_{RMSx}^2 + \sigma_{RMSy}^2}$.

Table 1

Combined standard deviation of the center of mass location in the intensity profile of the atomic columns in both x and y direction, obtained with different scanning methods and dwell times for a total of 966 atom columns per image.

| Dwell time/pixel (μs) | Raster scanning $\sigma_{RMS}$ (pixel) | Snake scanning corrected $\sigma_{RMS}$ (pixel) | Snake scanning $\sigma_{RMS}$ (pixel) | Hilbert scanning corrected $\sigma_{RMS}$ (pixel) | Hilbert scanning $\sigma_{RMS}$ (pixel) |
|---|---|---|---|---|---|
| 2 | 0.176 ± 0.002 | 0.173 ± 0.002 | 0.222 ± 0.003 | 0.160 ± 0.002 | 0.197 ± 0.002 |
| 4 | 0.140 ± 0.001 | 0.151 ± 0.002 | 0.151 ± 0.002 | 0.132 ± 0.001 | 0.165 ± 0.002 |
| 7 | 0.133 ± 0.001 | 0.129 ± 0.002 | 0.143 ± 0.002 | 0.111 ± 0.001 | 0.131 ± 0.001 |
| 8 | 0.134 ± 0.001 | 0.130 ± 0.002 | 0.164 ± 0.002 | 0.120 ± 0.001 | 0.130 ± 0.001 |
| 9 | 0.126 ± 0.001 | 0.126 ± 0.002 | 0.161 ± 0.002 | 0.116 ± 0.001 | 0.127 ± 0.001 |
| 10 | 0.134 ± 0.001 | 0.126 ± 0.001 | 0.164 ± 0.002 | 0.116 ± 0.001 | 0.118 ± 0.001 |
| 12 | 0.127 ± 0.001 | 0.123 ± 0.001 | 0.168 ± 0.002 | 0.111 ± 0.001 | 0.120 ± 0.001 |
| 15 | 0.123 ± 0.001 | 0.123 ± 0.001 | 0.176 ± 0.002 | 0.102 ± 0.001 | 0.121 ± 0.001 |
| 20 | 0.121 ± 0.001 | 0.115 ± 0.001 | 0.184 ± 0.002 | 0.100 ± 0.001 | 0.128 ± 0.001 |

Before looking at the quantification of the lattice distortions we focus again on the calculated diffractograms (Fourier transform, FT) of the images as these contain important information about the distortions and offer an intuitive way to describe them. As shown in the calculated FTs of the images acquired with the different scanning methods, insets in Figures 2(a1), 2(b1) and 2(c1), cross like features are identified over the spots that represent the spatial frequencies corresponding to the lattice. Commonly those features are associated with edge effects of the FT. To avoid those effects being mistaken as or associated with scanning distortions, before calculating the FT, a circular window with Gaussian smoothed edges ($\sigma$ = 10 pixels) was applied to the images acquired at 15 μs dwell time. The calculated FTs in Figure 4(a) and Figure 4(b) correspond to the images acquired with the raster and snake scanning method, respectively, after applying a circular window. The horizontal streaks over the spots of the FTs disappeared which was not the case for the vertical streaks. The FT in Figure 4(c) corresponds to the image acquired with the Hilbert scanning method, contrary to the previous two cases, applying a circular window removed both horizontal and vertical streaks around the spots of the FT.

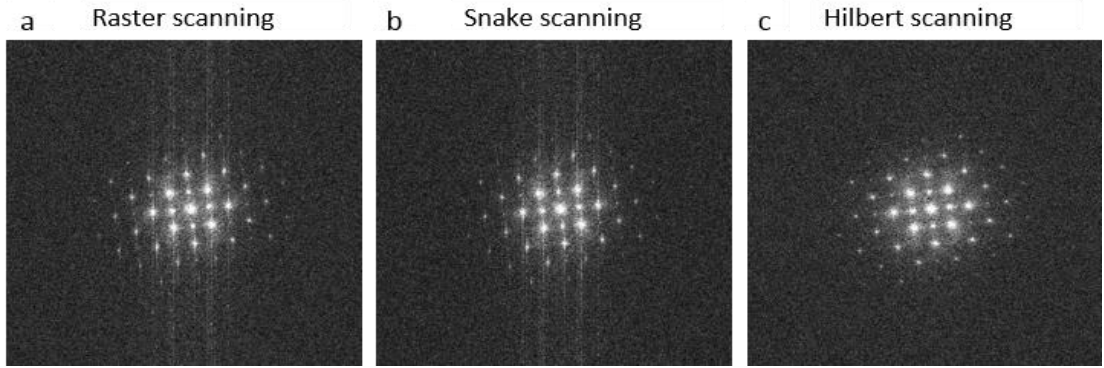

Figure 4. (a), (b) and (c), calculated FTs of the images acquired at a dwell time of 15 μs with the raster, snake and Hilbert scanning methods, respectively. A circular window with smooth edges was applied to the corresponding images to reduce edge effects in the FT. Spots on the FT of the same images without applying the circular window, shown in the insets of Figures 2(a1), 2(b1) and 2(c1), present cross like features. Note the absence of streaks in the Hilbert case.

Image distortions in the slow scan direction are clearly reduced when using Hilbert scanning, as vertical streaks in the calculated FT were not apparent. However, drift of the sample/stage is still present and results in abrupt discontinuities in the image where regions of the image that were taken with a longer time between them meet. This effect increases with increased dwell time, indicating that indeed this issue is related to slow sample drift rather than settling time issues with the scan system. Figure 5 shows these abrupt changes in images acquired at different dwell times. Indeed, rather than a line modulation as in raster and snake scanning, the modulation due to sample drift is now smeared over the image with approximately equal weight in all directions, leading to a better behaviour when calculating the diffractogram.

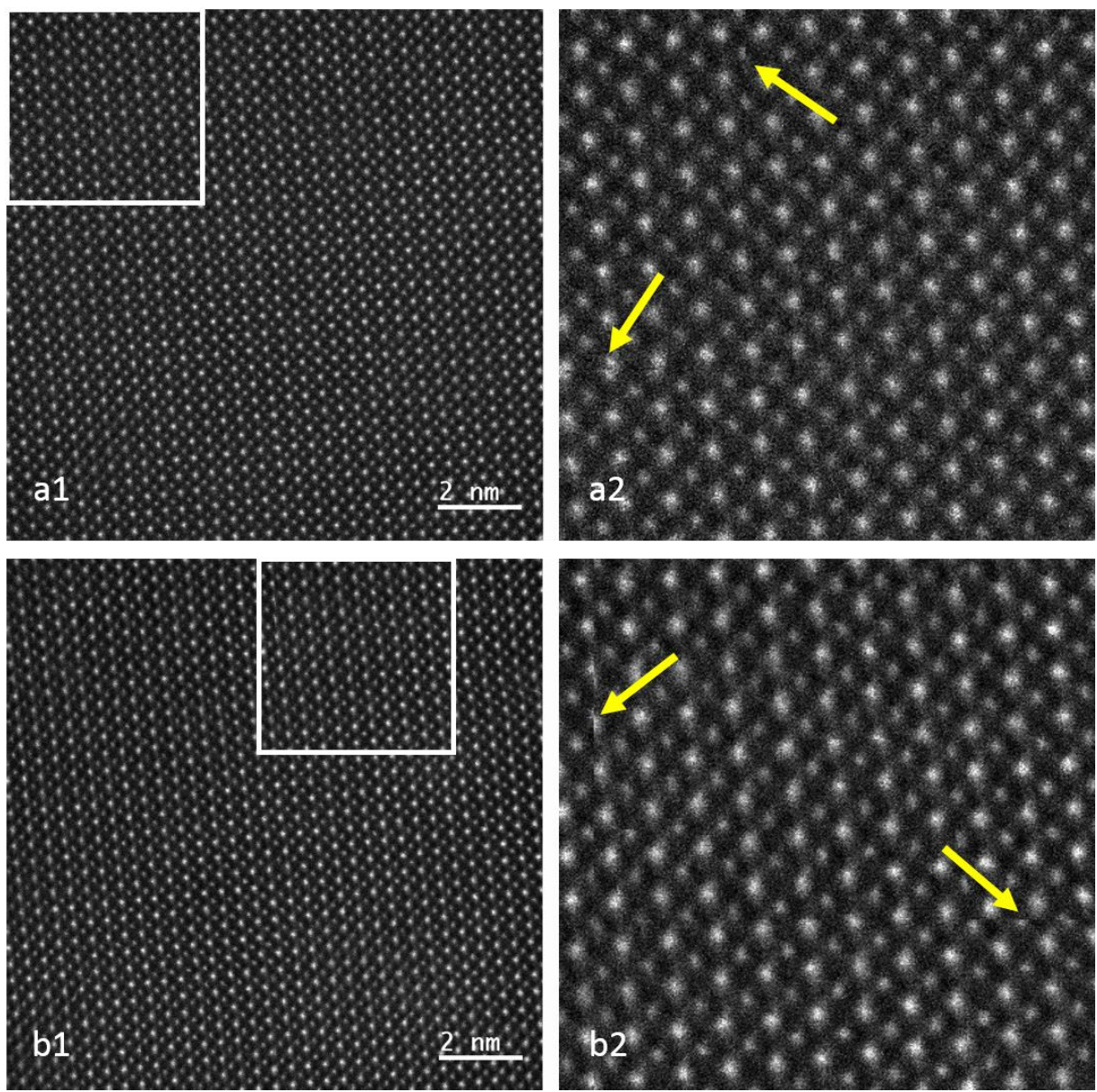

Figure 5. Experimental HAADF images. (a1) Image acquired with Hilbert scanning at 7 μs dwell time and (b1) at 20 μs dwell time. (a2) and (b2) are highlighted areas in (a1) and (b1), respectively. Yellow arrows indicate abrupt discontinuities in the images caused by the interplay between sample drift and the scan pattern.

Atomic resolution images contain direct information about strain, which can be very relevant in many material systems [18]. STEM images are however often problematic for this application as the sequential scanning mixes actual strain with artefacts caused by sample drift during the recording time. Strain mapping techniques offers the possibility to quantify these distortions [19, 20, 21, 22], in order to verify the role of the scan strategy for this application we decided to use strain mapping techniques based on real space analysis. We employed the Atomap software to calculate deviations in the interplanar spacing of the Sr sublattice in both scan directions. Because of the orientation of the sample with respect to the raster scanning directions, measuring the interatomic distances of the (010) planes will quantify the distortions generated in the slow scan direction and the interatomic distances of the (100) planes will quantify the distortions generated in the fast scan direction. In Figure 6 the interplanar spacing maps for the images acquired with the raster, snake and Hilbert scanning at 9 μs dwell time are shown. As the shape of the atomic columns does not significantly influence the interplanar spacing of the sublattices, for the results in Figures 6 we employed the corrected snake and Hilbert images instead of the uncorrected images.

The interplanar spacing maps of the (010) planes calculated from the images acquired with the raster and snake methods, Figure 6(a1) and Figure 6(b1), respectively, show horizontal bands commonly found because of drift leading to distortions in the slow scan direction. For the Hilbert case, the map is more homogeneous, Figure 6(c1). No bands are present in the corresponding maps of the (100) planes.

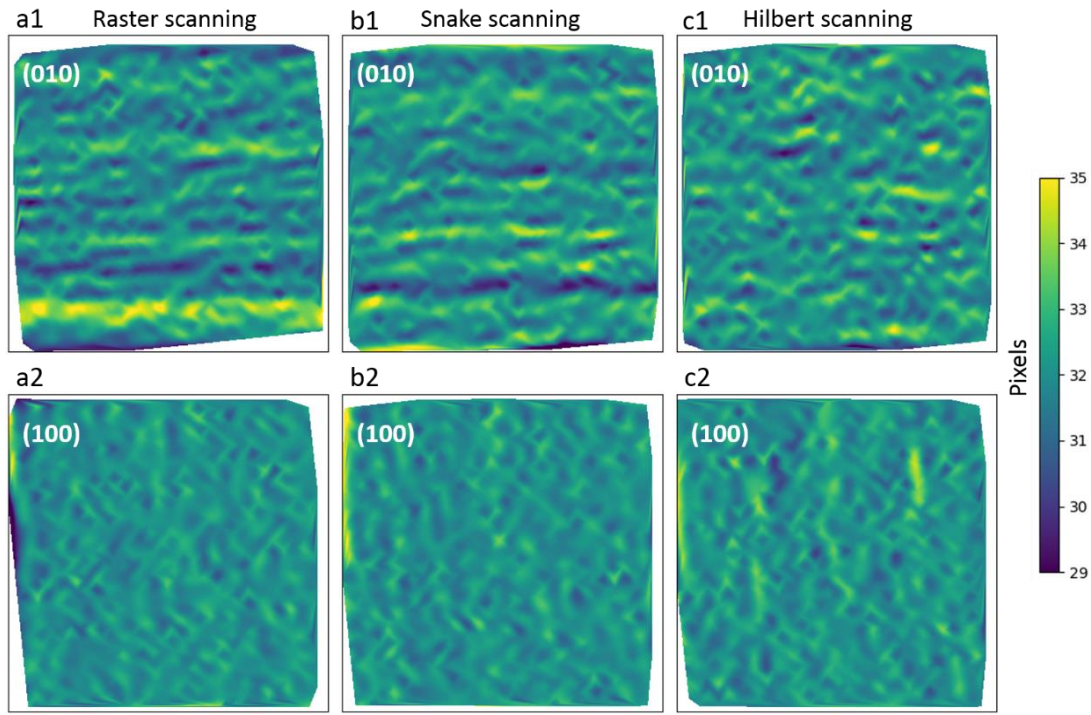

Figure 6. Interplanar spacing (010) and (100) of the Sr sublattice calculated from images with 9 µs dwell time and for the three different scanning patterns.

We plot the standard deviation in the interplanar spacing maps, such as the map in Figure 6, as a function of dwell time. The standard deviation of the interplanar spacing maps corresponding to the (010) planes are plotted in Figure 7(a) and the corresponding to the (100) planes are plotted in Figure 7(b). Here the results from the snake and Hilbert uncorrected and corrected images are included. It is not surprising that only slight differences between the results of the uncorrected and corrected data were found as there is not a substantial influence of the high frequencies distortions on the interplanar spacing. It can be seen that scanning with the Hilbert method at high speed acquisitions, 2 µs dwell time, generates more distortions than at slower speeds (these distortions were reduced for the corrected Hilbert images). The calculated FTs of the experimental Hilbert image acquired at 2 µs dwell time reveal extra spots at high frequency and half the Nyquist frequency and some of them are still present in the corrected Hilbert image (the experimental images acquired at 2 µs dwell time and their corresponding calculated FTs can be found in the Supporting Information Figure s5, extra spots in the FTs reveal the distortions at high speed acquisitions). As explained before, the continuous change of the scanning direction prevents the probe to reach a constant speed during the entire scanning path which is not the case for the raster scanning. At longer dwell times, when scanning with the Hilbert method the distortions become approximately constant for both scanning directions, while for the raster and snake methods the distortion first decrease and then keep increasing for the slow scan direction and first decreases and then becomes constant for the fast scan direction. This behaviour can be explained as follows. For very short dwell times, distortions occur due to the finite settling time of the probe on the sample. This occurs in this instrument and at the given magnification up to approximately 2 µs. Increasing the dwell time results in the disappearing of such artefacts, but now slow drift variations come into play that become more apparent with higher dwell times. This effect of drift depends strongly on the scan pattern and is low in the fast scan directions for raster and snake scanning, while it is high for the slow scan direction. In Hilbert scanning, there is no slow and fast axis and here the standard deviation is equally distributed in both directions with higher deviations for higher dwell times, to a small degree as expected (Figure 7, (a) and (b)). In Figure 7(c) the RMS of the standard deviations and the mean Sr interplanar spacing for the raster and corrected Hilbert scanning data are shown. The total distortions with the Hilbert method are comparable to the distortions obtained with the raster method.

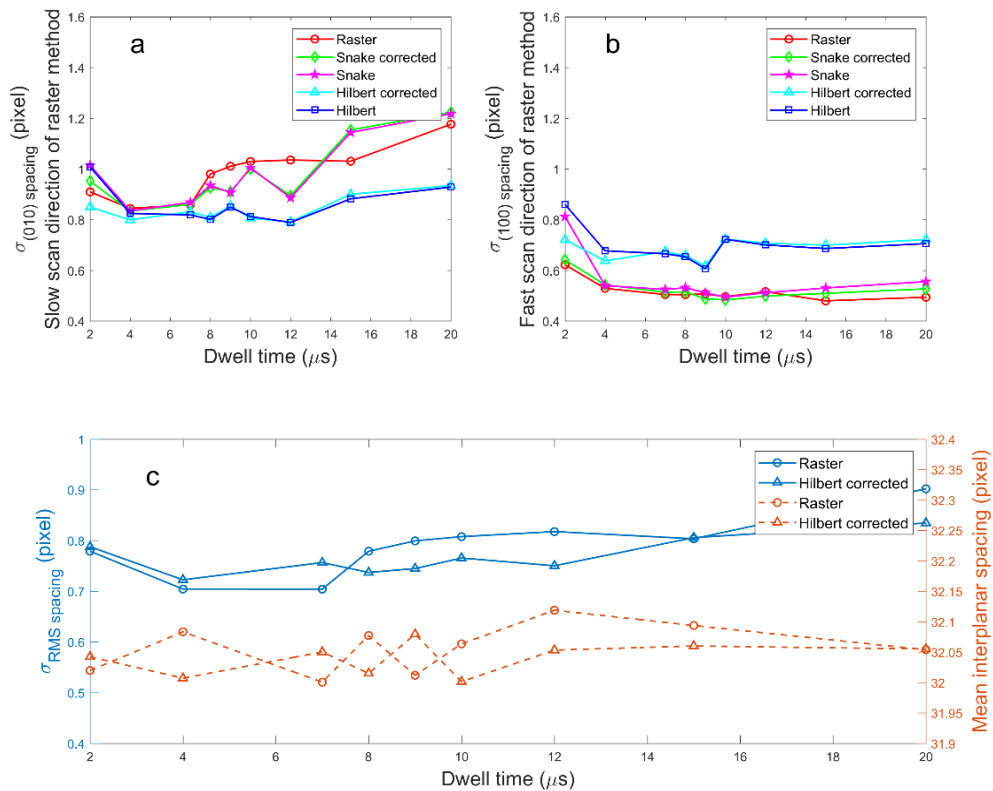

Figure 7. Standard deviations calculated from the interplanar spacing maps obtained from the images acquired with different scanning methods. (a) Standard deviations from the (010) planes, which quantifies the distortions in the slow scan direction of the raster method. (b) Standard deviations from the (100) planes, which quantifies the distortions in the fast scan direction of the raster method. (c) RMS of the standard deviations in both x and y and the mean interplanar spacing in x and y at different dwell times. This shows that Hilbert and raster scanning provide similar total standard deviations of the lattice spacing with no apparent bias for equal dwell times.

## 4. Discussions

Comparing the different scan regimes resulted in a few general observations. When it comes to the local high frequency information in an image, e.g. the shape of the atomic columns, we find that both raster and snake scanning introduce significant anisotropy because of the difference in fast and slow scan directions. This results in the presence of streaks in the diffractograms of the images. These streaks can be understood as a slow modulation affecting the different scan lines, and is caused by sample or probe drift on the sample. This issue can be alleviated with Hilbert scanning which shows a reduced total standard deviation around a localised feature with isotropic behaviour in both directions. This results in a much 'cleaner' diffractogram where the streak features are completely absent. In the image however, specific abrupt features appear along lines where pixels that are scanned with large time interval meet.

With regards to strain mapping, the scan pattern has an important effect due to drift leading to distortions in the image. For both raster and snake, this effect is very pronounced in the slow scan direction and for this reason, often two consecutive images are recorded with rotating the fast scan direction over 90 degrees.

For the Hilbert scan, the distortion effect due to drift is also more isotropic while the total precision of the strain measurement depends entirely on the amount of drift and the total acquisition time as long as very short acquisition times are avoided. The total standard deviation caused by drift in terms of strain mapping is however nearly independent of the scan method and one could see this is as an 'error-budget' that is more equally distributed over x and y direction for Hilbert scanning as opposed to the two other scan methods.

There are however two important benefits in doing a single Hilbert scan as compared to two orthogonal raster or snake scans.

(i) Single scan is faster and requires a lower dose, which might be essential for beam sensitive samples.

(ii) A single scan containing correct information about x and y strain allows to correctly calculate the relation between both strain directions which is important for e.g. shear strain. This is more difficult for raster scanning as it requires combining low noise x and y strain maps from two different images.

In terms of dose, both snake and Hilbert scanning are preferred as no flyback time is required (unless a fast beam blanker is available in which case the raster scanning will provide the same dose). The distribution of the dose with the Hilbert method is clearly different than with the raster method. The Hilbert method registers small neighbouring areas of the sample in a shorter time than line by line scanning, this could have an influence on beam damage as the dose is distributed faster in smaller areas, further experiments need to be performed on beam sensitive samples to clarify these effects. In the field of selective laser sintering, it was shown that the distribution of temperature in the treated object was more uniform when scanning with a Hilbert method compared to a raster method [23].

The Hilbert space filling curve of order n subdivides a square grid in $4^n$ sub-squares, for this reason only frame sizes with equal dimensions of a power of 2 are feasible but as these are very common dimensions in raster scanning too, this is not an important restriction.

## 5. Conclusions

An alternative rectangular scanning method was proposed to reduce distortions and flyback delay in STEM. At moderate scanning speeds, the Hilbert scanning method showed comparable image quality and distortions with respect to the conventional raster scanning. No post-processing was required to achieve the results other than implementing an experimentally determined time lag representing the finite response of the system. At these conditions the common vertical scanning distortions in the slow scan direction caused by drift were reduced which is obvious from the shape of sharp features in the images, the diffractograms and when performing real space strain mapping. We showed that for strain mapping the total noise budget remains the same over the different scan strategies, but Hilbert scan distributes this noise budget equally over the scan directions which avoids the need for two consecutive perpendicular scans as is commonly used with raster scanning.

As the method can be implemented on any existing STEM instrument, either by the manufacturer changing the scan engine firmware/hardware or alternatively by the customer adding an external scan engine, it could find rapid uptake in the community.

**Acknowledgements**

A.V., A.B. and J.V. acknowledge funding through FWO project G093417N ('Compressed sensing enabling low dose imaging in transmission electron microscopy') from the Flanders Research Fund. M.N. received support for this work from the European Union's Horizon 2020 research and innovation programme under the Marie Skłodowska-Curie grant agreement No 838001. J.V acknowledges funding from the European Union's Horizon 2020 research and innovation programme under grant agreement No 823717 – ESTEEM3.

**References**

[1]  S.J. Pennycook, M.F. Chisholm, A.R. Lupini, M. Varela, A.Y. Borisevich, M.P. Oxley, W.D. Luo, K. Van Benthem, S.H. Oh, D.L. Sales, S.I. Molina, J. GarcíA-Barriocanal, C. Leon, J. SantamaríA, S.N. Rashkeev, S.T. Pantelides, Aberration-corrected scanning transmission electron microscopy: From atomic imaging and analysis to solving energy problems, Philos. Trans. R. Soc. A Math. Phys. Eng. Sci. 367 (2009) 3709–3733. https://doi.org/10.1098/rsta.2009.0112.

[2]  P.E. Batson, N. Dellby, O.L. Krivanek, Sub-ångstrom resolution using aberration corrected electron optics, Nature. 418 (2002) 617–620. https://doi.org/10.1038/nature00972.

[3]  H.S. von Harrach, Instrumental factors in high-resolution FEG STEM, Ultramicroscopy. 58 (1995) 1–5. https://doi.org/10.1016/0304-3991(94)00172-J.

[4]  D.A. Muller, E.J. Kirkland, M.G. Thomas, J.L. Grazul, L. Fitting, M. Weyland, Room design for high-performance electron microscopy, Ultramicroscopy. 106 (2006) 1033–1040. https://doi.org/10.1016/j.ultramic.2006.04.017.

[5]  A. Muller, J. Grazul, Optimizing the environment for sub-0.2 nm scanning transmission electron microscopy, J. Electron Microsc. (Tokyo). 50 (2001) 219–226. https://doi.org/10.1093/jmicro/50.3.219.

[6]  S. Ning, T. Fujita, A. Nie, Z. Wang, X. Xu, J. Chen, M. Chen, S. Yao, T.-Y. Zhang, Scanning distortion correction in STEM images, Ultramicroscopy. 184 (2018) 274–283. https://doi.org/https://doi.org/10.1016/j.ultramic.2017.09.003.

[7]  L. Jones, P.D. Nellist, Identifying and correcting scan noise and drift in the scanning transmission electron microscope, Microsc. Microanal. 19 (2013) 1050–1060. https://doi.org/10.1017/S1431927613001402.

[8]  J.P. Buban, Q. Ramasse, B. Gipson, N.D. Browning, H. Stahlberg, High-resolution low-dose scanning transmission electron microscopy, J. Electron Microsc. (Tokyo). 59 (2010) 103–112. https://doi.org/10.1093/jmicro/dfp052.

[9]  C. Ophus, J. Ciston, C.T. Nelson, Correcting nonlinear drift distortion of scanning probe and scanning transmission electron microscopies from image pairs with orthogonal scan directions, Ultramicroscopy. 162 (2016) 1–9. https://doi.org/10.1016/j.ultramic.2015.12.002.

[10]  L. Jones, H. Yang, T.J. Pennycook, M.S.J. Marshall, S. Van Aert, N.D. Browning, M.R. Castell, P.D. Nellist, Smart Align—a new tool for robust non-rigid registration of scanning microscope data, Adv. Struct. Chem. Imaging. 1 (2015) 1–16. https://doi.org/10.1186/s40679-015-0008-4.

[11]  X. Sang, A.R. Lupini, R.R. Unocic, M. Chi, A.Y. Borisevich, S. V Kalinin, E. Endeve, R.K. Archibald, S. Jesse, Dynamic scan control in STEM: spiral scans, Adv. Struct. Chem. Imaging. 2 (2016) 6. https://doi.org/10.1186/s40679-016-0020-3.

[12]  L. Kovarik, A. Stevens, A. Liyu, N.D. Browning, Implementing an accurate and rapid sparse sampling approach for low-dose atomic resolution STEM imaging, Appl. Phys. Lett. 109 (2016). https://doi.org/10.1063/1.4965720.

[13]  D. Hilbert, Ueber die stetige Abbildung einer Linie auf ein Flähenstück. (On Continuous Mapping of a Line onto a Planar Surface), Math. Ann. 38 (1891) 459–460.

[14]  P. Klapetek, A. Yacoot, P. Grolich, M. Valtr, D. Nečas, Gwyscan: a library to support non-equidistant scanning probe microscope measurements, Meas. Sci. Technol. 28 (2017) 34015. https://doi.org/10.1088/1361-6501/28/3/034015.

[15]  M. Tence, M. Kociak. Collaboration for Custom scanning hardware.

[16]  W.C. Lenthe, J.C. Stinville, M.P. Echlin, Z. Chen, S. Daly, T.M. Pollock, Advanced detector signal acquisition and electron beam scanning for high resolution SEM imaging, Ultramicroscopy. 195 (2018) 93–100. https://doi.org/10.1016/j.ultramic.2018.08.025.

[17]  M. Nord, P.E. Vullum, I. MacLaren, T. Tybell, R. Holmestad, Atomap: a new software tool for the automated analysis of atomic resolution images using two-dimensional Gaussian fitting, Adv. Struct. Chem. Imaging. 3 (2017). https://doi.org/10.1186/s40679-017-0042-5.

[18]  V. Prabhakara, D. Jannis, A. Béché, H. Bender, J. Verbeeck, Strain measurement in semiconductor FinFET devices using a novel moiré demodulation technique, Semicond. Sci. Technol. (2019). http://iopscience.iop.org/10.1088/1361-6641/ab5da2.

[19]  A.M. Sanchez, P.L. Galindo, S. Kret, M. Falke, R. Beanland, P.J. Goodhew, An approach to the systematic distortion correction in aberration-corrected HAADF images, J. Microsc. 221 (2006) 1–7. https://doi.org/10.1111/j.1365-2818.2006.01533.x.

[20]  L. Jones, S. Wenner, M. Nord, P.H. Ninive, O.M. Løvvik, R. Holmestad, P.D. Nellist, Optimising multi-frame ADF-STEM for high-precision atomic-resolution strain mapping, Ultramicroscopy. 179 (2017) 57–62. https://doi.org/10.1016/j.ultramic.2017.04.007.

[21]  G. Bárcena-González, M.P. Guerrero-Lebrero, E. Guerrero, D. Fernández-Reyes, D. González, A. Mayoral, A.D. Utrilla, J.M. Ulloa, P.L. Galindo, Strain mapping accuracy improvement using super-resolution techniques, J. Microsc. 262 (2016) 50–58. https://doi.org/10.1111/jmi.12341.

[22]  J. Li, S. Cheng, L. Wu, J. Tao, Y. Zhu, The effect of scanning jitter on geometric phase analysis in STEM images, Ultramicroscopy. 194 (2018) 167–174. https://doi.org/10.1016/j.ultramic.2018.07.011.

[23]  L. Ma, H. Bin, Temperature and stress analysis and simulation in fractal scanning-based laser sintering, Int. J. Adv. Manuf. Technol. 34 (2007) 898–903. https://doi.org/10.1007/s00170-006-0665-5.

**Supporting information**

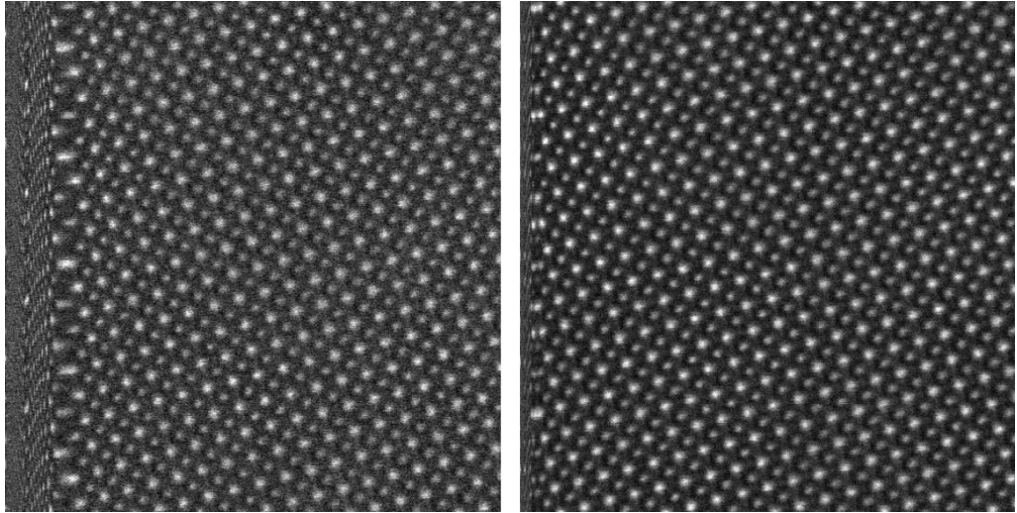

Figure s1. Experimental images acquired with the raster method without any flyback delay. Left, image acquired at 1 µs dwell time. Right, image acquired at 4 µs dwell time. There are clear distortions on the left side of the images, which are reduced when the dwell time is increased.

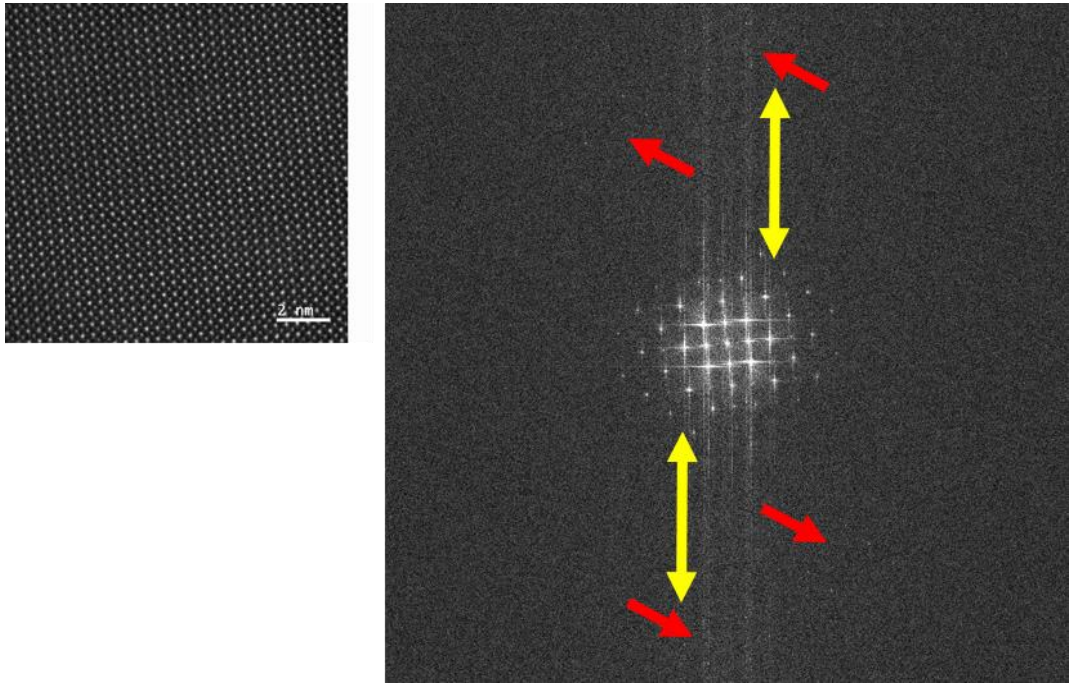

Figure s2. Left, experimental image acquired with the raster method with 10 μs dwell time. Right, its corresponding calculated FT. Vertical streaks are indicated by yellow double arrows, and faint extra spots are indicated by red single arrows.

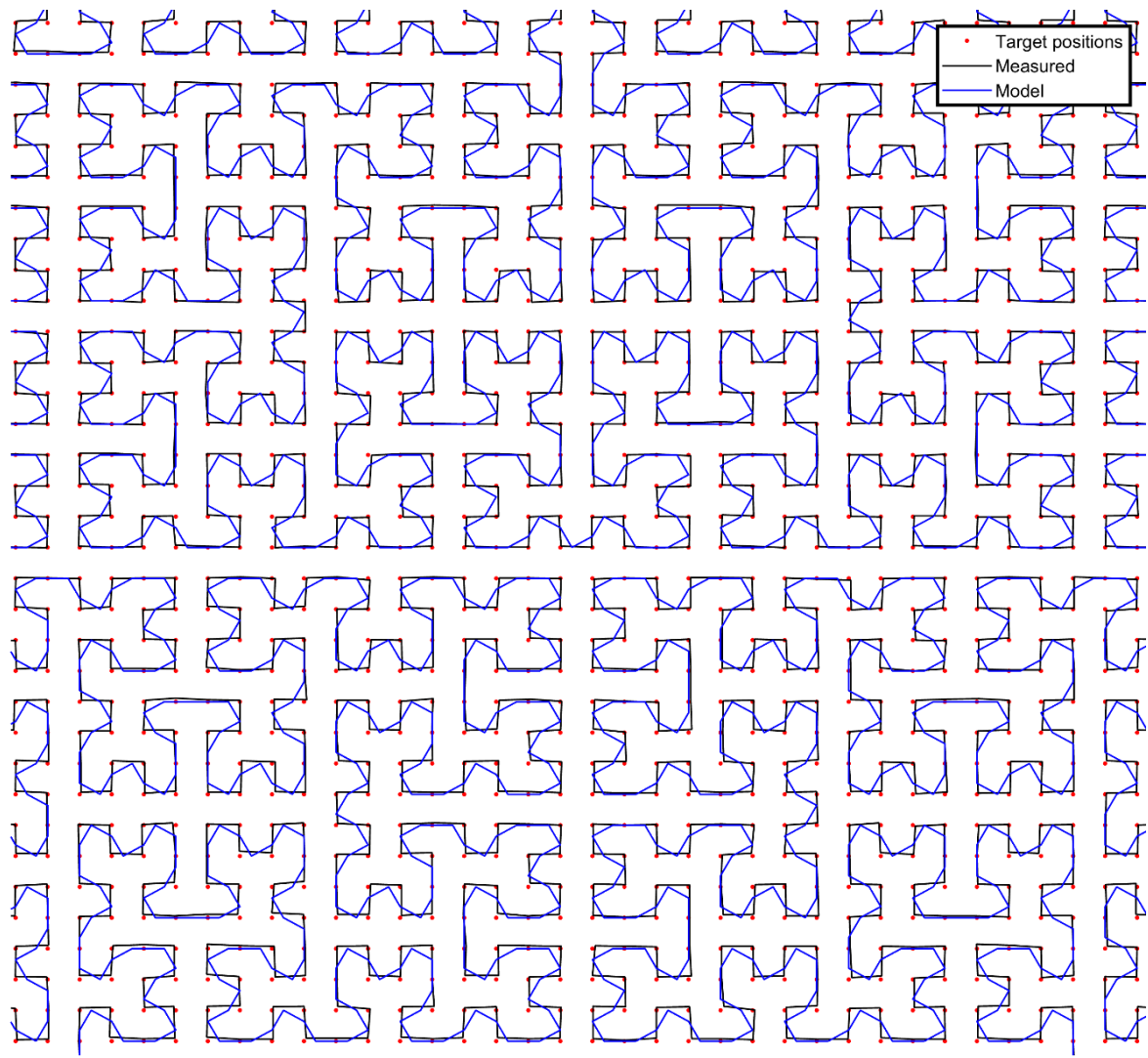

Figure s3. Target positions, the scanning path as measured from the scan generator and the model used to correct the Hilbert image acquired at 15 μs dwell time.

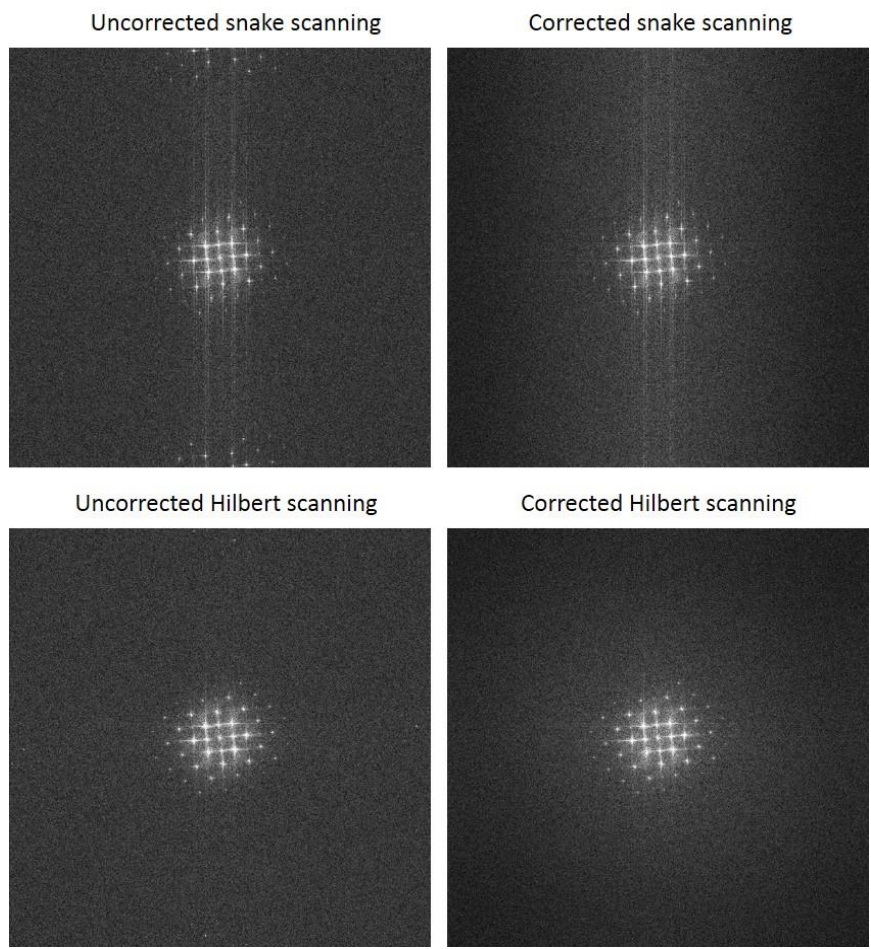

Figure s4. Left, calculated FT of the images acquired with the snake and Hilbert method at 15 μs dwell time. Right, calculated FT of the corresponding corrected images.

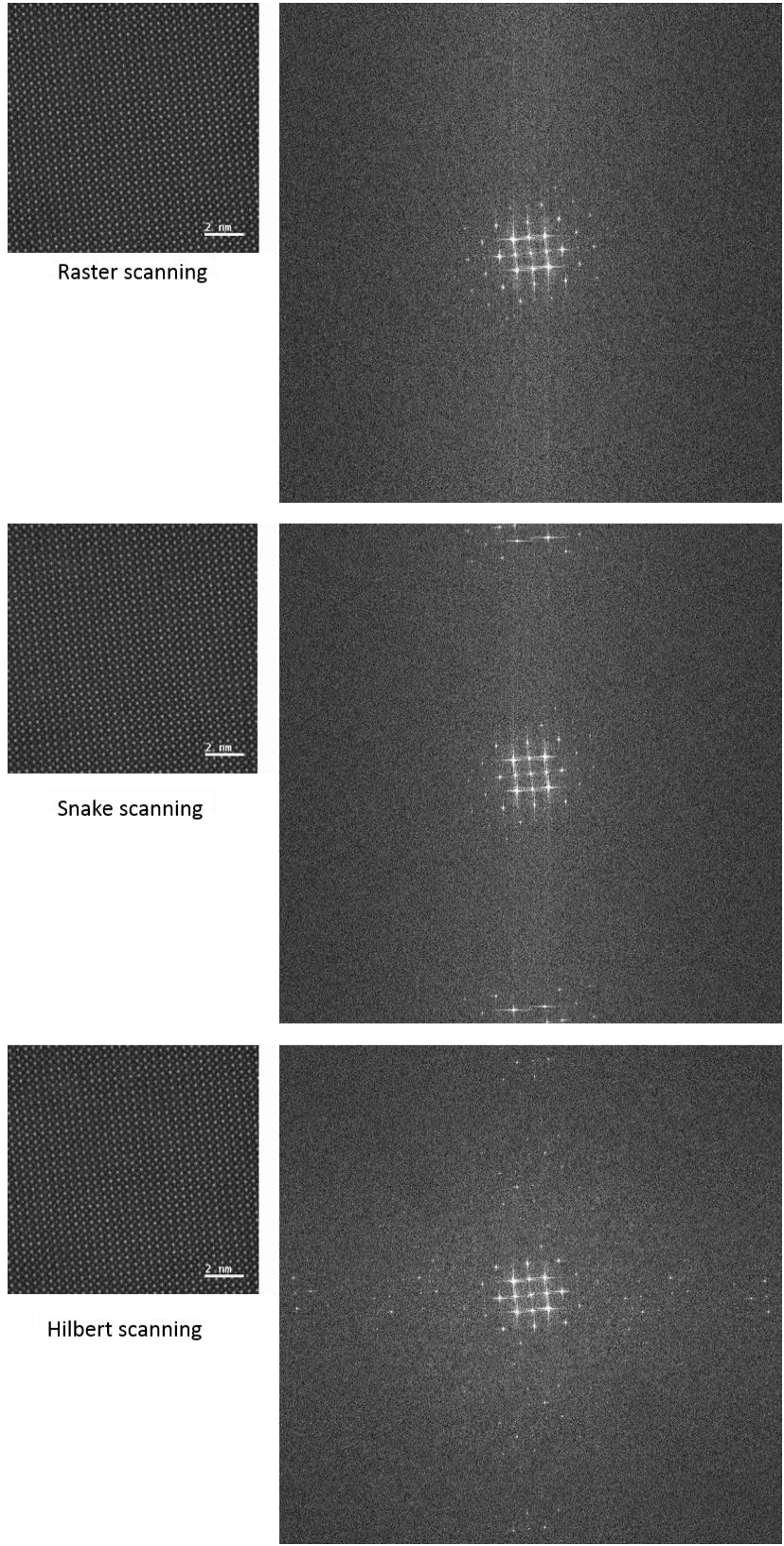

Figure s5. Experimental images acquired with the raster, snake and Hilbert method at 2 μs dwell time and their corresponding calculated FTs.